\documentclass[onecolumn,showpacs]{revtex4}
\pdfoutput=1

\usepackage{graphicx}%
\usepackage{dcolumn}
\usepackage{amsmath}

\makeatletter
\def\btt#1{\texttt{\@backslashchar#1}}%
\DeclareRobustCommand\bblash{\btt{\@backslashchar}}%
\makeatother

\begin{document}


\title{The Effect of Student Learning Styles on the Learning Gains Achieved When Interactive Simulations Are Coupled with Real-Time Formative Assessment via Pen-Enabled Mobile Technology }

\author{F.V. Kowalski and S.E. Kowalski}
\affiliation{Physics Department, Colorado
School of Mines, Golden CO. 80401 U.S.A.}

\begin{abstract}
This paper describes results from a project in an undergraduate engineering physics course that coupled classroom use of interactive computer simulations with the collection of real-time formative assessment using pen-enabled mobile technology.  Interactive simulations (free or textbook-based) are widely used across the undergraduate science and engineering curriculia to help actively engaged students increase their understanding of abstract concepts or phenomena which are not directly or easily observable.  However, there are indications in the literature that we do not yet know the pedagogical best practices associated with their use to maximize learning.  This project couples student use of interactive simulations with the gathering of real-time formative assessment via pen-enabled mobile technology (in this case, Tablet PCs).  The research question addressed in this paper is:  are learning gains achieved with this coupled model greater for certain types of learners in undergraduate STEM classrooms?  To answer this, we correlate learning gains with various learning styles, as identified using the Index of Learning Styles (ILS) developed by Felder and Soloman. These insights will be useful for others who use interactive computer simulations in their instruction and other adopters of this pedagogical model; the insights may have broader implications about modification of instruction to address various learning styles.
\end{abstract}


\maketitle

\section{Introduction}
\label{sec:intro}

Explorations of new pedagogical models have many facets and produce rich insights into how science and engineering education can be enhanced and invigorated.  In a two-semester study, we have been investigating the effectiveness of coupling two paradigms based on active learning strategies and constructivist learning theory: using computer simulations and gathering real-time formative assessment. To accomplish this, each student used a single, pen-enabled mobile device. Elsewhere, we describe emerging evidence of the effectiveness of this coupling in undergraduate engineering physics \cite{kowalski} and chemical engineering \cite{gardner} courses.

However, the intriguing tangential question we consider here is:  are the learning gains achieved when this model is implemented greater for STEM students of particular learning styles?  It could be hypothesized, for example, that this learning activity, based on a visually-rich experience and hands-on manipulation of  variables, is more clearly aligned with some learning style preferences and therefore might be more successful for certain students (e.g., visual, sensing, and active learners) than others.

\section{BRIEF DESCRIPTION OF THE COUPLED MODEL}
\label{sec:nurturing}

The coupling of interactive simulations with real-time formative assessment will hereafter be referred to as ``the Coupled Model."  In this case, the coupling was implemented using one Tablet PC per student, but other types of  increasingly available and increasingly affordable pen-enabled mobile devices could accomplish this as well.  Thus the coupled model is economically feasible using devices provided by the institution or the students themselves.

Many STEM educators use interactive simulations (a.k.a. ``sims," ``applets," etc.) to help actively engaged students better understand abstract concepts or phenomena which are not directly or easily observable.  There are abundant free, high-quality, web-based simulations available online, even for advanced topics, and others are available in association with textbooks.  However, as we examine pedagogical best practices for implementing the use of these to enhance learning, shortcomings have been identified with teacher manipulation of the simulations \cite{lane}, with independent student exploration of the simulations \cite{delmas}, and with more precise (``cookbook") direction of student exploration \cite{wieman}.  It seems that students need skillful guidance from the instructor as they use interactive simulations to discover insights, and careful monitoring as they draw conclusions about new principles and abstract concepts they investigate.

Both of these needs can be met by gathering real-time formative assessment. This cornerstone of constructivist learning is most effective when students respond to open-format (vs. multiple choice) questions, as higher levels of thinking can better be probed and richer insights into student thinking can be achieved. The questions asked for this assessment provide a vehicle for actively engaging students with the course content in the context of the interactive simulation, optimizing use of class time, and increasing student metacognition.  Perhaps most importantly in the Coupled Model, these questions can guide learning through timely scaffolded questioning and feedback, helping students to attain a more mature understanding of difficult concepts.

Previously, gathering real-time formative assessment to open-format questions on a time scale that made it meaningful was a daunting task \cite{black}.  Socrates did it successfully with a small group of students, but most contemporary STEM instructors do not have that luxury.   However, we have developed and maintain InkSurvey. This free, web-based software allows students equipped with pen-enabled mobile technology to use digital ink to respond with words, diagrams, graphs, equations, proofs, etc. to open-format questions posed by the instructor.  The instructor, receiving these responses instantaneously, has unprecedented real-time insight into student thinking and can reinforce correct understandings and modify misconceptions on the most immediate of time scales.  There is other software, both free and commercially available, that will facilitate this, but we chose InkSurvey since it is designed specifically for this purpose and has been used successfully in classes with enrollments exceeding 60 students.   An additional feature of this use of technology is that a single device in the hands of the student can be used for both the interactive simulation and the construction of the response; this adds to its pedagogical appeal \cite{young}.

\section{LEARNING STYLES }
\label{sec:styles}

Over the past few decades, the study of human learning styles has emerged as a well-established field within the discipline of cognitive psychology.  Although many different assessments of student learning styles have been used effectively in investigating how STEM concepts and skills are learned [e.g., \cite{mccaulley,sharp,heywood}], we utilize the widely accepted \cite{litzinger} Index of Learning Styles (ILS) developed by Felder and Soloman \cite{felder1,felder2}. This instrument's roots are in engineering education and it is available for free online \cite{index}.  This inventory and the awareness of various student learning styles that it stimulated impacted engineering education. Indeed, by 2000, leaders in engineering education considered identifying and teaching for various learning styles to be of great importance in improving the classroom environment in undergraduate engineering classes \cite{klinger}.
Based on responses to 44 items in the ILS questionnaire (each one forcing the students to choose between two alternatives), every student's learning style can be placed somewhere along each of four continua.  These reflect the preferred way the student perceives information (sensory vs. intuitive learning style), how sensory information is most effectively perceived (visual vs. verbal learning style), how the student processes information (active vs. reflective learning style), and how the student progresses toward understanding (sequential vs. global learning style).

Previous studies indicate that certain learning activities are more effective for engineering students with particular learning styles. For example, when a process control course in electrical engineering was redesigned to introduce active, collaborative learning and to increase student exposure to real-life problems, including the use of advanced computer simulations, researchers found significantly greater learning gains were achieved by students with ``active" and ``global" learning styles \cite{zywno1}.

\section{PROCEDURE}
\label{sec:procedure}

The goal of the course targeted for this study,  ``Advanced Physics Laboratory I," is to introduce students to the process of modeling in engineering and science. This is illustrated predominately using wave phenomena but also requires coverage of fundamental statistical content. The latter is introduced in the first three weeks of the course via a one week laboratory exercise along with reading, and homework assignments from the text \cite{baird}. Much of the content delivery in class is associated with computer simulations coupled with real-time formative assessment (Coupled Model).

The students enrolled in the two sections are engineering physics juniors having 3 semesters each of physics and calculus, along with courses in differential equations, linear algebra, thermodynamics, analog electronics, and a summer session in vacuum systems, optics, machine shop, computer interfacing, and electronics.

The research presented here is focused on how student understanding of content, using the Coupled Model, is influenced by learning style. For measuring learning gains, a pre-test at the beginning of the semester and a summative evaluation at the end of the semester were administered, thus determining learning of content presented via the coupled model.

Simulations involving phasors in circuits, Fourier analysis, and diffraction, along with two others in statistics, were used. The procedure for measuring learning in the phasor circuits simulation is illustrated next. Details for the others will appear in another publication.

\underline{Learning objective}: to understand how phasors are used in applying Kirchhoff's voltage law to a circuit by summing time-varying sinusoidal voltages of the same frequency across the circuit elements.

\underline{Pre-test Question}: ``How are phasors used in circuit analysis?"

Students were asked to play before class with the simulation found here:\\*
\url{http://vnatsci.ltu.edu/s_schneider/physlets/main/phasor1.shtml}

\underline{Post-play InkSurvey question used in class}: ``What does this simulation illustrate?"

\underline{Scaffolding InkSurvey question used in class}: ``What did you observe in the circuit simulation relative to your understanding of the phasors?"

\underline{Summative assessment question}: ``Explain how phasors are used in the analysis of circuits."

Student responses were then evaluated based on their demonstration of understanding that Kirchhoff's law is satisfied at each instant of time, using the rubric in Table I.

\begin{table}[ht]
\caption{SCORING RUBRIC FOR  PHASOR CIRCUITS SIMULATION}
\centering
\begin{tabular}{p{5 cm}|p{5 cm}|p{5 cm}}
\hline\hline
Unsatisfactory & Satisfactory & Excellent \\[0.5ex]
\hline
No discussion of this relationship & Discusses Kirchhoff's law in a static phasor diagram & Relates Kirchhoff's law via a phasor diagram for different times \\
Score=0 & Score=1 & Score=2 \\ [1ex]
\hline
\end{tabular}
\label{table:rubric}
\end{table}

\section{Results}
\label{sec:results}

\subsection{Student Misconceptions Revealed by InkSurvey}

As noted above, all students were asked to run the simulation and to explain (via InkSurvey) what they learned from the interactive simulation relative to their understanding of circuits. Typically, students did not notice that at each point in time on the animation, the voltage across the source equals the sum of the voltages across the inductor and resistor. The instructor, reading the earliest submissions on InkSurvey, appropriately coaxed them to think about Kirchhoff's law and how that relates to the simulation. Students revised scaffolding InkSurvey submissions as necessary.

\begin{table}[ht]
\caption{LEARNING GAINS AND CORRELATION COEFFICIENTS FOR LEARNING STYLES AND LEARNING GAINS ACHIEVED, FOR FIVE SIMULATIONS.  LEARNING GAIN FOR EACH SIMULATION IS CLASS AVERAGE (SEE TEXT).  TABLE SUMMARIZES  PEARSON R VALUE AND ACCEPTABLE RANGE FOR 95\% CONFIDENCE.}
\centering
\begin{tabular}{|c|p{0.5 cm}|c|c|c|c|c|}
\hline
Simulation & N & Ave. gain & ACT-REF & SEN-INT & VIS-VRB & SEQ-GLO \\[0.5ex]
\hline
1. Circuits & 41 & 63\% & 0.02 &  0.02 & -0.04 & -0.12 \\
\hline
\multicolumn{3}{|c|}{95\% confidence range} & -0.28 to 0.32 & -0.28 to 0.32 & -0.34 to 0.26 & -0.41 to 0.19\\ [1ex]
\hline
2. Fourier & 41 & 67\% & 0.03 &  -0.004 & -0.06 & -0.05 \\
\hline
\multicolumn{3}{|c|}{95\% confidence range} & -0.27 to 0.33 & -0.30 to 0.30 & -0.35 to 0.25 & -0.35 to 0.25\\ [1ex]
\hline
3. Diffraction & 30 & 60\% & 0.03 &  0.01 & 0.10 & -0.15 \\
\hline
\multicolumn{3}{|c|}{95\% confidence range} & -0.32 to 0.37 & -0.34 to 0.36 & -0.26 to 0.43 & -0.47 to 0.21\\ [1ex]
\hline
4. Std. Dev. Mean & 40 & 75\% & 0.25 &  0.04 & 0.14 & -0.11 \\
\hline
\multicolumn{3}{|c|}{95\% confidence range} & -0.06 to 0.51 & -0.27 to 0.34 & -0.17 to 0.43 & -0.40 to 0.20\\ [1ex]
\hline
5. Least Squares & 33 & 61\% & -0.13 &  -0.28 & 0.03 &0.06 \\
\hline
\multicolumn{3}{|c|}{95\% confidence range} & -0.44 to 0.21 & -0.56 to 0.06 & -0.30 to 0.36 & -0.28 to 0.39\\ [1ex]
\hline
\end{tabular}
\label{table:results}
\end{table}

\subsection{Analysis of data}

All assessments were identically scaled with an appropriate rubric. Next, learning gains were calculated as average normalized gains, specifically the difference between summative and pre-test assessments divided by the difference between a perfect score and the pre-test assessment.   The class average learning gain for each simulation is presented in Table II. Data from students who did not attend the coupled model class session were removed from this analysis. The p values for paired two tailed t-test results with unequal variance, all less than $0.001$, showed significant learning gains which were most likely not due to chance. The same conclusion (every $p < 0.001$) was reached using the Kolmogorov-Smirnov test.

To determine the effect of learning styles (dependent variable) on learning gains (independent variable) achieved using the coupled model, we measured learning styles using the ILS learning style inventory \cite{index}. These styles are categorized in complementary pairs. A score is given for each element of the pair. For example, the instrument measures active and reflective styles (ACT-REF) each on a scale of $0-11$. The other measures are sensing and intuitive (SEN-INT), visual and verbal (VIS-VRB), and sequential and global (SEQ-GLO) learning styles. FIG. 1 shows the learning styles distribution for the class, offered in two sections, with a total of $41$ students.

A negative scale is introduced here to more easily graph the results. Data indicating a score on the first characteristic in the learning pair is shown as negative with a more negative value indicating a stronger learning style for the first member of the pair while the positive axis records the other aspect of the learning style pair.

\begin{center}
\begin{figure}
\includegraphics[scale=0.6]{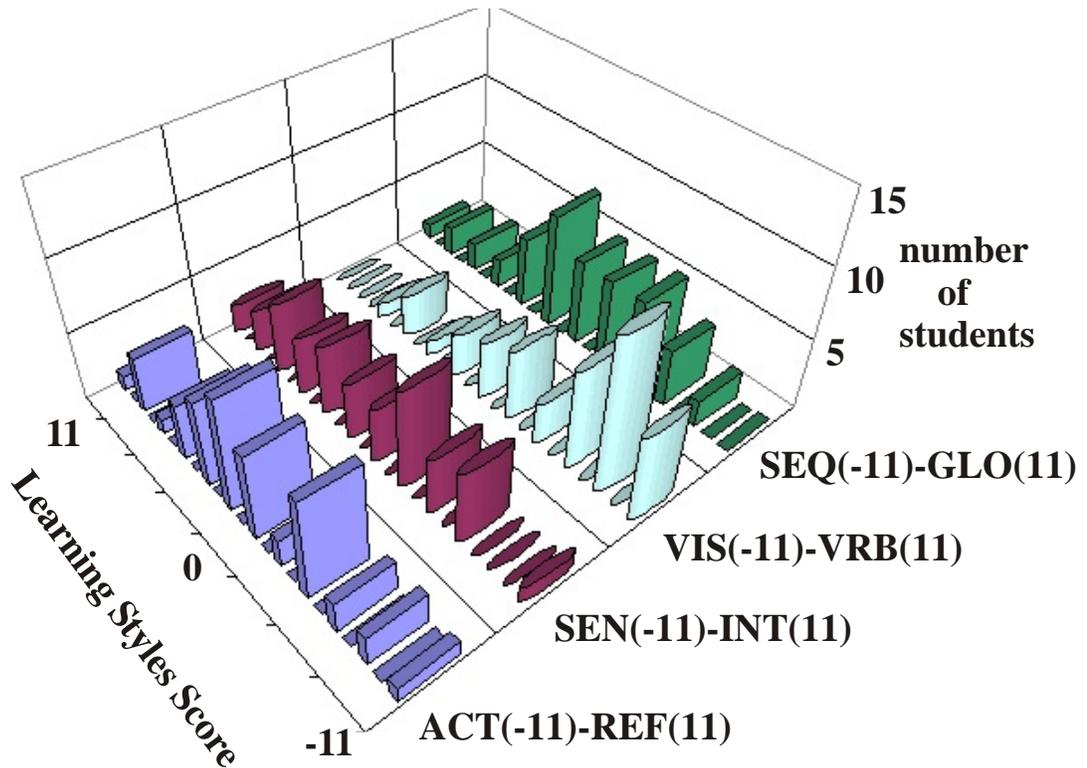}
\caption{Learning Style Scores vs. Number of Students.}
\label{fig:learningstyles}
\end{figure}
\end{center}

For example, in the VIS-VRB pair, ``visual" is the first term so a score further along the negative axis indicates more of a visual learning style, while further along the positive axis indicates more of a verbal style.	
The independent variable, learning gain, was then correlated using a Pearson r value with each dependent variable corresponding to a particular ILS learning style and shown in Table II. The significance of each correlation coefficient is reported in Table II as a range within which the population correlation coefficient lies with $95\%$ confidence \cite{altman}.
Reliability and construct validity of the ILS learning styles measurement has been documented \cite{litzinger,felder3,zywno2}. To test reliability of the learning gains measurement in this study, nine assessments (both pre-test and summative evaluations) were measured by four different raters. The inter-rater reliability was calculated based on the work of Ebel \cite{ebel} using web-based software \cite{solomon}. The resulting nine reliability estimates were then averaged to yield a value $0.82$ with a standard deviation 0.12. The reliability estimate ranges between zero and one with the larger value indicating more reliability between raters.	 

\subsection{Discussion}

Although the original hypothesis at the root of this study was that the coupling of interactive computer simulations and real-time formative assessment would better meet the learning styles of some students than others, this is not supported by the learning gains data.  As shown in Table II, this coupling resulted in significant learning gains for all categories of student learning styles; there is no statistically significant correlation between student learning style and the effectiveness of this pedagogical method in producing learning gains in this population of engineering physics students.

It is interesting to note that the three simulations on which students achieved the greatest learning gains (circuits, Fourier, and standard deviation mean) were also the three simulations with the greatest number of students present in class.  We do not know the reasons for diminished attendance on certain dates, but speculate that the student population present on any given day is not entirely random.  This may account for the observed differences in learning gains achieved; however, the data suggest it is not a function of student learning styles.

The distribution of the learning styles of the engineering physics students included in this study did not closely mirror that previously reported in the literature for electrical \cite{zywno1,odwyer}, chemical \cite{montgomery}, bioengineering and mechanical \cite{patten}, general \cite{rosati} engineering undergraduate students, and others.  Although all of these groups shared a preponderance of visual and sequential learning styles, the engineering physics students in this study had a greater population of students with reflective and intuitive learning styles than those consistently reported elsewhere for groups of engineering undergraduates. Nevertheless, even though the experimental group may not precisely reflect the learning styles of other known populations of engineering students, they represent a broad distribution of learning styles, as appropriate for this study.

It is greatly encouraging that this pedagogical model (coupling interactive computer simulations and real-time formative assessment) is not preferentially advantageous to any particular learning style, but rather is broadly effective across all learning styles.

\section{Conclusions}
\label{sec:conclusions}

When interactive computer simulations were coupled with  real-time formative assessment collected with InkSurvey, students achieved large and statistically significant learning gains on all five examples. These learning gains, however, are shown to have low correlation with student learning styles. The sample population used in this study has been measured to contain a broad distribution of learning styles. The reliability of the instrument to measure student learning styles has been documented, and the inter-rater reliability on measurement of student learning gains is very strong. We therefore interpret the low correlation between learning gains and learning style to indicate that the Coupled Model is effective in enhancing learning, independent of student learning style, in an engineering physics classroom.

\begin{acknowledgments}
This material is based upon work supported by a 2011 HP Catalyst Grant and by the National Science Foundation under Grant \# 1037519.
\end{acknowledgments}


\begin{references}


\bibitem{kowalski} F.V. Kowalski, and S.E. Kowalski, ``Enhancing curiosity using interactive simulations combined with real-time formative assessment facilitated by open-format questions on tablet computers,"  Frontiers in Education Conference, Seattle, WA. Oct. 2012.

\bibitem{gardner} T.Q. Gardner,  F.V. Kowalski, and S.E. Kowalski, ``Interactive simulations coupled with real-time formative assessment to enhance student learning,"  Proceedings of the 2012 ASEE Conference and Exposition, San Antonio, TX. June 2012.

\bibitem{lane} D.M. Lane and S.C. Peres, ``Interactive simulations in the teaching of statistics: promise and pitfalls," International Conference on Teaching Statistics (ICOTS)-7, 2006.

\bibitem{delmas} R. delMas, J. Garfield, and B. Chance, ``A model of classroom research in action: developing simulation activities to improve students' statistical reasoning," Journal of Statistics Education, 7(3), 1999.

\bibitem{wieman} C.E. Wieman, K.K. Perkins, and W.K. Adams, ``Interactive simulations for teaching physics:  what works, what doesn't, and why,"  American Journal of Physics, 76 (4 \& 5), pp. 393-399, 2007.

\bibitem{black} P. Black, and D. Wiliam, ``Inside the black box: raising standards through classroom assessment."  Phi Delta Kappan, 80 (2), pp. 139-148, 1998.

\bibitem{young} M. F. Young, ``Assessment of situated learning using computer environments," Journal of Science Education and Technology, Vol. 4(1), pp. 89-96, 1995.

\bibitem{mccaulley} M.H. McCaulley, E.S. Godleski, C.F. Yokomoto, L. Harrisberger, and E.D. Sloan,  ``Applications of psychological type in engineering education,"  Engineering Education,  pp. 394-400, Feb. 1983.

\bibitem{sharp} J.E. Sharp, ``Teaching teamwork communication with Kolb learning style theory," Proceedings of the 31st ASEE/IEEE Frontiers in Education Conference, Reno NV,  Oct. 2001.

\bibitem{heywood} J. Heywood, J., Engineering Education: Research and Development in Curriculum and Instruction, Hoboken, New Jersey: IEEE Press, John Wiley \& Sons, Inc.,  pp. 122-123, 2005.

\bibitem{litzinger} T.A. Litzinger, S.H. Lee, J.C. Wise, and R.M.Felder, ``A psychometric study of the index of learning styles,"  Journal of Engineering Education, 96 (4), pp. 309-319, 2007.

\bibitem{felder1} R.M. Felder and L.K. Silverman, ``Learning styles and teaching styles in engineering education,"  Engineering Education, 78(7), pp. 674-681, 1988.

\bibitem{felder2} R.M Felder, ``Reaching the second tier:  learning and teaching styles in college science education,"  Journal of College Science Teaching, 23(5), pp. 286-290, 1993


\bibitem{index} Index of Learning Styles Questionnaire, available online at \url{http://www.engr.ncsu.edu/learningstyles/ilsweb.html}

\bibitem{klinger} A. Klinger, C.J. Finelli, \& D.D. Budny, ``Improving the classroom environment,"  Proceedings of the 30th ASEE/IEEE Frontiers in Education Conference,  2000.

\bibitem{zywno1} M.S. Zywno, and J.K. Waalen,  ``The Effect of hypermedia instruction on  achievement and attitudes of students with different learning styles,"  Proceedings of the 2001 ASEE Conference and Exposition, 2001.

\bibitem{baird} D.C. Baird, {\bf Experimentation: An Introduction to Measurement Theory and Experiment Design}, Englewood Cliffs, NJ: Prentice-Hall, Inc., 1995.

\bibitem{altman} D.G. Altman, D. Machin, T.N. Bryant, and M.J. Gardner, Statistics with Confidence, Second Edition, BMJ Books, ISBN 0 7279 1375 1. p. 89-92, 2000.

\bibitem{felder3} R.M. Felder and J. Spurlin,  ``Applications, reliability, and validity of the Index of Learning Styles," International Journal of Engineering Education, 21 (1), pp.103-112, 2005.

\bibitem{zywno2} M.S. Zywno, ``A contribution to validation of score meaning for Felder-Soloman's Index of Learning Styles," Proceedings of the 2003 ASEE Conference and Exposition, 2003.

\bibitem{ebel} R. Ebel, ``Estimation of the reliability of ratings," Psychometrika, 16, pp. 407-424, 1951.

\bibitem{solomon} D.J. Solomon, ``The rating reliability calculator." BMC Medical Research Methodology,  4:11, 2004.

\bibitem{odwyer} A. O"Dwyer, A., ``Learning styles of first year level 7 electrical and mechanical engineering students at DIT,"  Proceedings of the 2008 International Symposium for Engineering Education, pp. 69-74, 2008.

\bibitem{montgomery} S. Montgomery, ``Addressing diverse student learning styles through the use of multimedia,"  Proceedings of the 1995 Frontiers in Education Conference, 1995.

\bibitem{patten} E. Patten,  S. Atwood, and L. Pruitt, ``Use of learning styles for teamwork and professional development in a multidisciplinary course,"  Proceedings of the 2010 ASEE Conference , 2010.

\bibitem{rosati} P. Rosati, ``Specific differences and Ssmilarities in the learning preferences of engineering students,"  Proceedings of the 1999 ASEE Conference and Exposition, 1999.



\end{references}
\end{document}